\documentclass[aps,reprint,showpacs,floatfix,pre,twocolumn]{revtex4-1}
\usepackage{amsmath}
\usepackage{graphicx}
\usepackage{amssymb}
\usepackage{epsfig}
\usepackage{epsf}
\usepackage{color}
\usepackage{fancyhdr}
\usepackage{float}
\newcommand\be {\begin{equation}}
\newcommand\ee {\end{equation}}
\newcommand\bea {\begin{eqnarray}}
\newcommand\eea {\end{eqnarray}}
\newcommand\n {\nonumber}
\newcommand\bc {\begin{center}}
\newcommand\ec {\end{center}}
\newcommand\bfl{\begin{flushleft}}
\newcommand\efl{\end{flushleft}}
\newcommand\bfr{\begin{flushright}}
\newcommand\efr{\end{flushright}}
\begin{document}
\title {Characterization of kinetic coarsening in a random field 
Ising model}
\author{Pradipta Kumar Mandal}
\email{pradipta.mandal@gmail.com}
\affiliation{Department of Physics, Scottish Church College,\\
 1 $\&$ 3 Urquhart Square Kolkata 700 006, India}
\author{Suman Sinha}
\email{suman.sinha.phys@gmail.com}
\affiliation{Department of Physics, University of Calcutta, \\
92 Acharya Prafulla Chandra Road, Kolkata 700009, India}
\begin{abstract}
We report a study of non equilibrium relaxation in a two-dimensional 
random field Ising model at a non-zero temperature. We attempt to observe 
the coarsening from a different perspective with a particular focus on 
three dynamical quantities that characterize the kinetic coarsening. We 
provide a simple generalised scaling relation of coarsening supported by 
numerical results. The excellent data collapse of the dynamical quantities 
justifies our proposition. The scaling relation corroborates the recent 
observation that the average linear domain size satisfies different scaling 
behaviour in different time regime.
\end{abstract}
\maketitle 
Study of the effect of disorder on non-disordered
magnetic systems has been a subject of intense interest for the last
several years \cite{im,gg,sra,aw,fv,cadmz,lkfi,sjk,puri,pp}. 
When a magnetic system is quenched from a high temperature to a
low temperature, it locally orders with the formation of domains
separated by domain walls. 
The average linear size of the domains
$R(t)$ grows with time. 
This linear size can also be understood as a
non equilibrium correlation length of the system. 
The growth of the
characteristic length scale $R(t)$ with time is known as the coarsening of
the system. Although coarsening in non-disordered systems is
well understood \cite{brayre}, progress in understanding the same in
disordered systems has been rather slow. 
Unavailability of reliable theoretical tools makes it
difficult to study the dynamics of disordered systems out of
equilibrium. Moreover, the dynamics of disordered systems is typically
so slow that we cannot access the truly asymptotic time regime in
numerical simulations. Despite all these, last several years have
witnessed an appreciable development in the study of disordered 
systems. These include coarsening of disordered magnets
\cite{acc,hp,lac}, polymers in random media \cite{krg,np,mg} or vortex
lines in disordered type - II superconductors \cite{nj,bcd,dla}.
The fundamental quantity of interest in the coarsening is the
growing length scale $R(t)$ and almost all studies of coarsening is
primarily concerned with the determination of this 
$R(t)$. However the growth law governing the coarsening
of disordered systems is at the center of some
controversies. Some numerical simulations on disordered ferromagnets
\cite{ppr1,ppr2,psr} yielded an algebraic growth $R(t) \sim t^{1/z}$, with a 
non universal dynamical exponent $z$ that depends on the temperature
and on the nature of disorder. Huse and Henley \cite{hh} suggested a
logarithmic increase of $R(t) \sim (\ln
t)^{1/\psi}$, with the barrier exponent $\psi > 0$. Later a series of
papers on the dynamics of elastic lines in a random potential
\cite{np,ibk,mg} claims a dynamic crossover from a pre-asymptotic
algebraic regime to a asymptotic slow logarithmic regime. 
Recent studies on other disordered systems \cite{puri2,puri3} 
supports this claim too.

In this work, we investigate the coarsening dynamics of a disordered
system, namely the random field Ising model (RFIM), focusing
our attention on three morphological
quantities which are functions of the strength of the
random fields ($\eta_0$) and the temperature ($T$). These are the
total length of the interfaces ($\Pi(\eta_0,t)$), i.e., the total
number of boundary spins of all the domains, the total number of domains
($\Lambda(\eta_0,t)$) in the system and the length of the interface of
the domain with largest mass ($\Omega(\eta_0,t)$), i.e., the number of
boundary spins of the domain containing maximum number of spins.
%The interest lies in the scaling and universality in a problem
%dominated by a complex energy landscape. 
In this work, we 
provide a empirical scaling relation of the coarsening.  
The scaling relation
is found to be nicely obeyed by the three morphological
quantities and is capable of explaining coarsening in disordered
magnets in the conventional way, i.e., the behaviour of the
average linear domain size for the entire time regime can be
reproduced from the proposed scaling relation and in this sense
it is general.
%show that the
%scaling relation is quite general. The scaling relation is
%able to demonstrate that in the initial time regime, the
%algebraic growth is only transient followed by a crossover to
%a slow logarithmic growth in late time regime. It has the 
%novelty of capturing all the salient features of the coarsening
%in a single framework and is thus of general importance in this
%area of research.

The Hamiltonian of the RFIM is given by
\be
\label{eqn1}
  H = - J\sum_{\langle i,j \rangle} s_i s_j + \sum_i \eta_i s_i +
  H_{\tt{ext}} \sum_i s_i
\ee
where $s_i = \pm 1$ is the spin variable at site $i$, $J$ is the strength
of the exchange interaction (conventionally set to unity) and $\eta_i$ is the
quenched random fields taken from an uniform distribution with varying 
strength $\eta_0$. The external field $H_{\tt {ext}}$ has been set to zero 
to observe the unbiased dynamics of the system. We consider a $L \times L$ 
square lattice (here $L=256$) with periodic boundary conditions along 
both directions. We start our simulations from a completely random spin 
configurations, characteristic of a high temperature ($T=\infty$) phase
and then suddenly quench the system to a temperature $T=0.50$, well 
below the critical temperature of a non-disordered system (Ising model) 
and then observe the time evolution of the system. The single spin flip
Metropolis algorithm \cite{metro} is used to simulate the system. 
Here an unit of time (ie, a time step) refers to one Monte Carlo (MC) step
and one MC step is taken to be completed (ie, $t=1$) when the number of
attempted single spin moves equals the total number of spins in the system.
The number 
of domains with their sizes are determined by the Hoshen-Kopelman 
algorithm \cite{hk}. All the quantities are averaged over $50$ independent
simulations to get a precise estimate. The quantities are normalized with 
respect to the total number of spins ($L^2$) of the system. The temperature
is taken sufficiently low to reduce the thermal fluctuations. 

The relevance of domain wall roughening due to temperature in comparison 
to that due to random fields is explained by Binder \cite{bind}.
For low temperature, the length scale ($\sim \text{exp} \left(2J/K_BT \right)$)
over which the thermal fluctuations is relevant is much higher than that 
($\sim \left(J/\eta_0\right)^2$) due to random fields fluctuations. 
A critical temperature may be obtained from the comparison of the length 
scales beyond which the effect of the temperature and the random fields 
on the domain wall roughening are significant. The dynamics of the system
at low temperature like $T=0.50$ will match that at $T=0$. Later we 
show that the time evolution of the three quantities introduced earlier 
is governed by the minimization of the total energy for the 
domain formation. Thus as long as $T$ is small, ground state (GS)
can be approached gradually as $t \to \infty$.
So after long time the final state is statistically the same as the
ground state (GS), i.e., the overlap between the GS and the
corresponding finite-$T$ state is close to unity for $T$ small.
As a whole, coarsening proceeds through a compromise
between the strength of the exchange interaction and the random fields
with the thermal fluctuations serving only to renormalize the
strengths of these couplings \cite{wm}.

\begin{figure}[h!]
  \centering
  \includegraphics[scale=0.30,angle=-90]{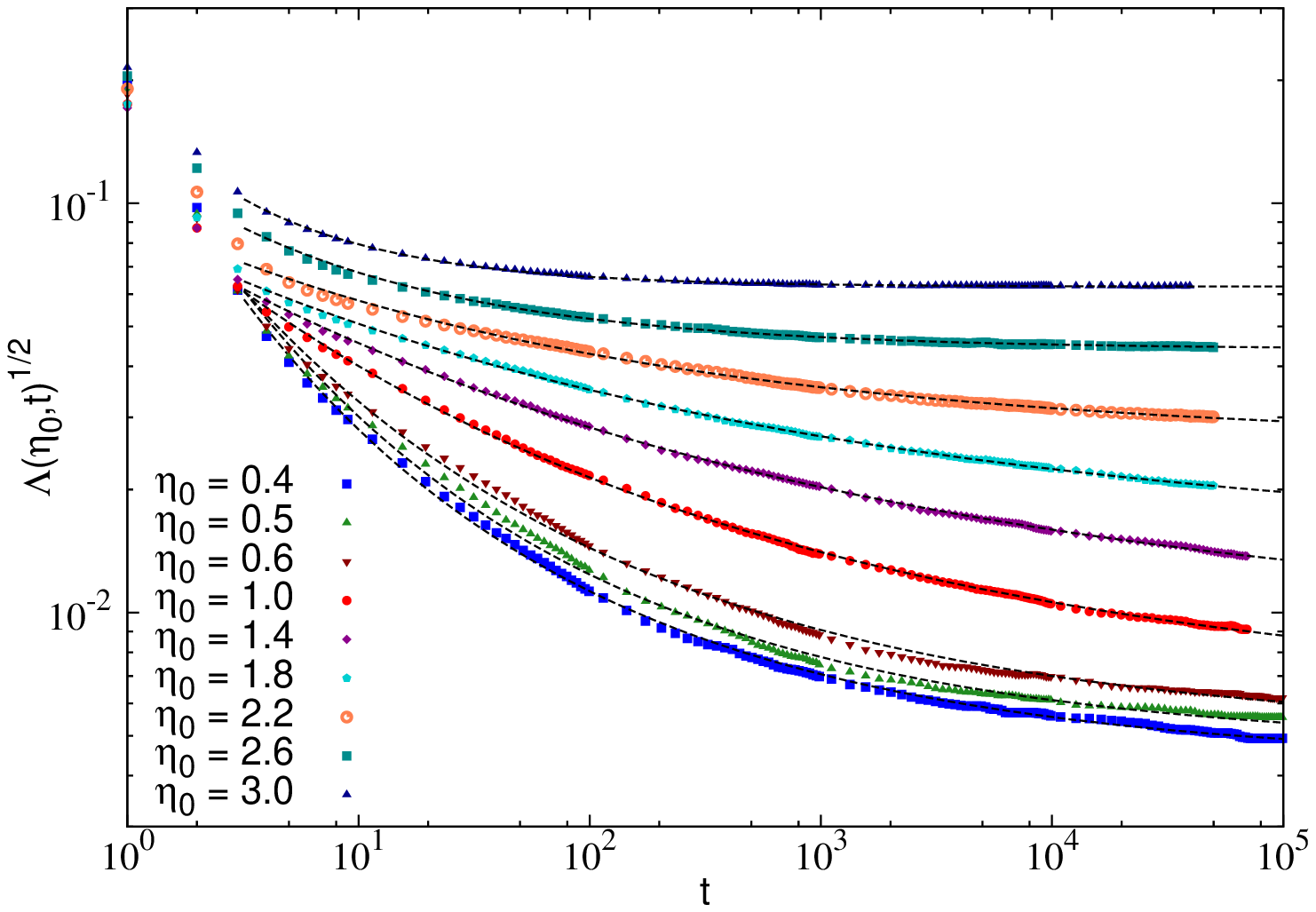}
  \caption{(Color online) The plot of the function $\Pi(\eta_0,t)$  
     against time. 
    The dotted lines are the best fits according to the scaling
    relation (\ref{eqn6}).}
  \label{fig1}
\end{figure}

\begin{figure}[h!]
  \centering
  \includegraphics[scale=0.30,angle=-90]{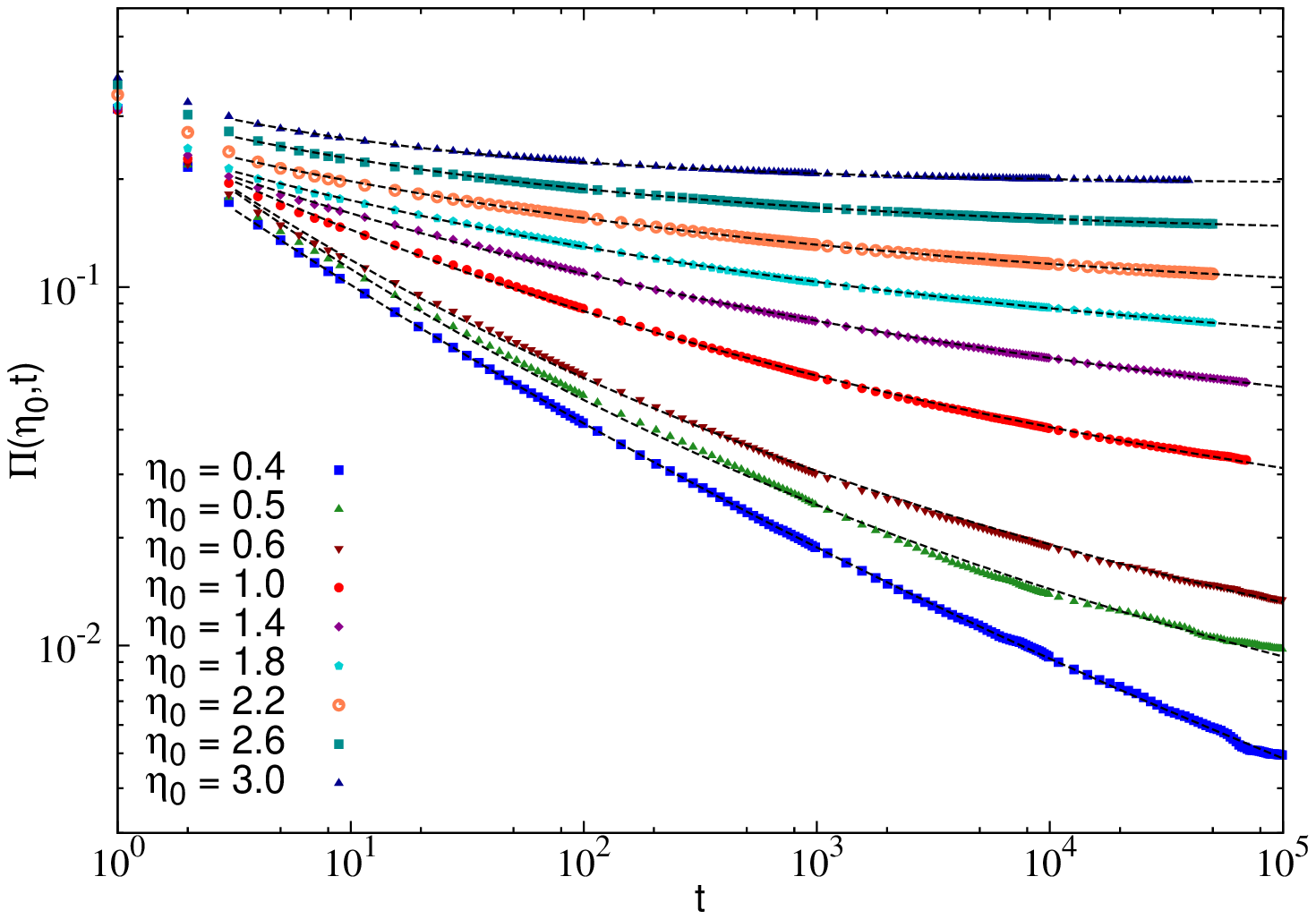}
  \caption{(Color online) The plot of the function  
    $\Lambda(\eta_0,t)^{1/2}$ against time. 
    The dotted lines are the best fits according to the scaling
    relation (\ref{eqn6}).}
  \label{fig2}
\end{figure}

\begin{figure}[h!]
    \centering
    \includegraphics[scale=0.30,angle=-90]{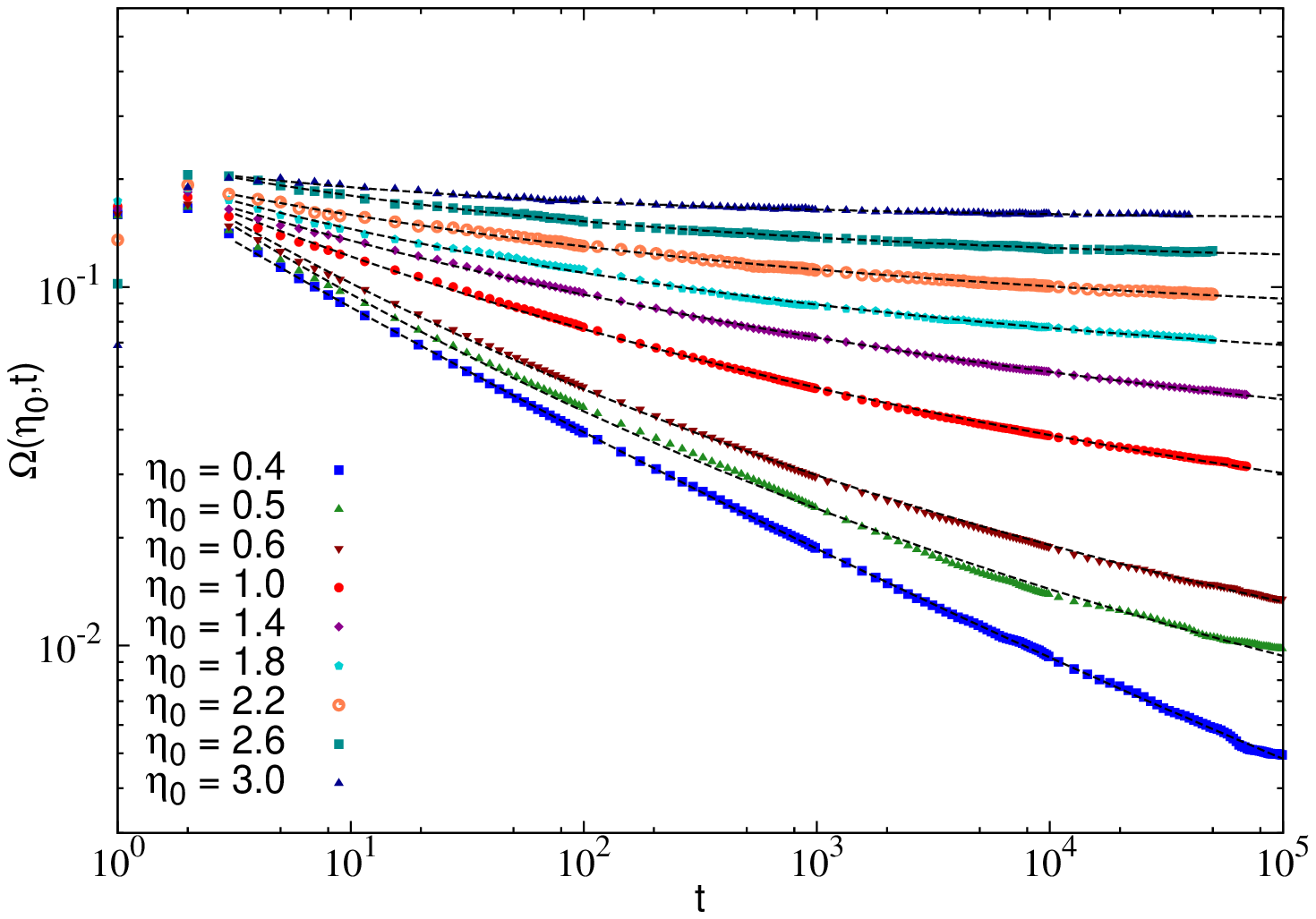}
    \caption{(Color online) Plot of $\Omega(\eta_0,t)$ against time along 
      with the
      best fits according to (\ref{eqn6}).}
    \label{fig3}
  \end{figure}

\begin{figure}[h!]
  \includegraphics[scale=0.30,angle=-90]{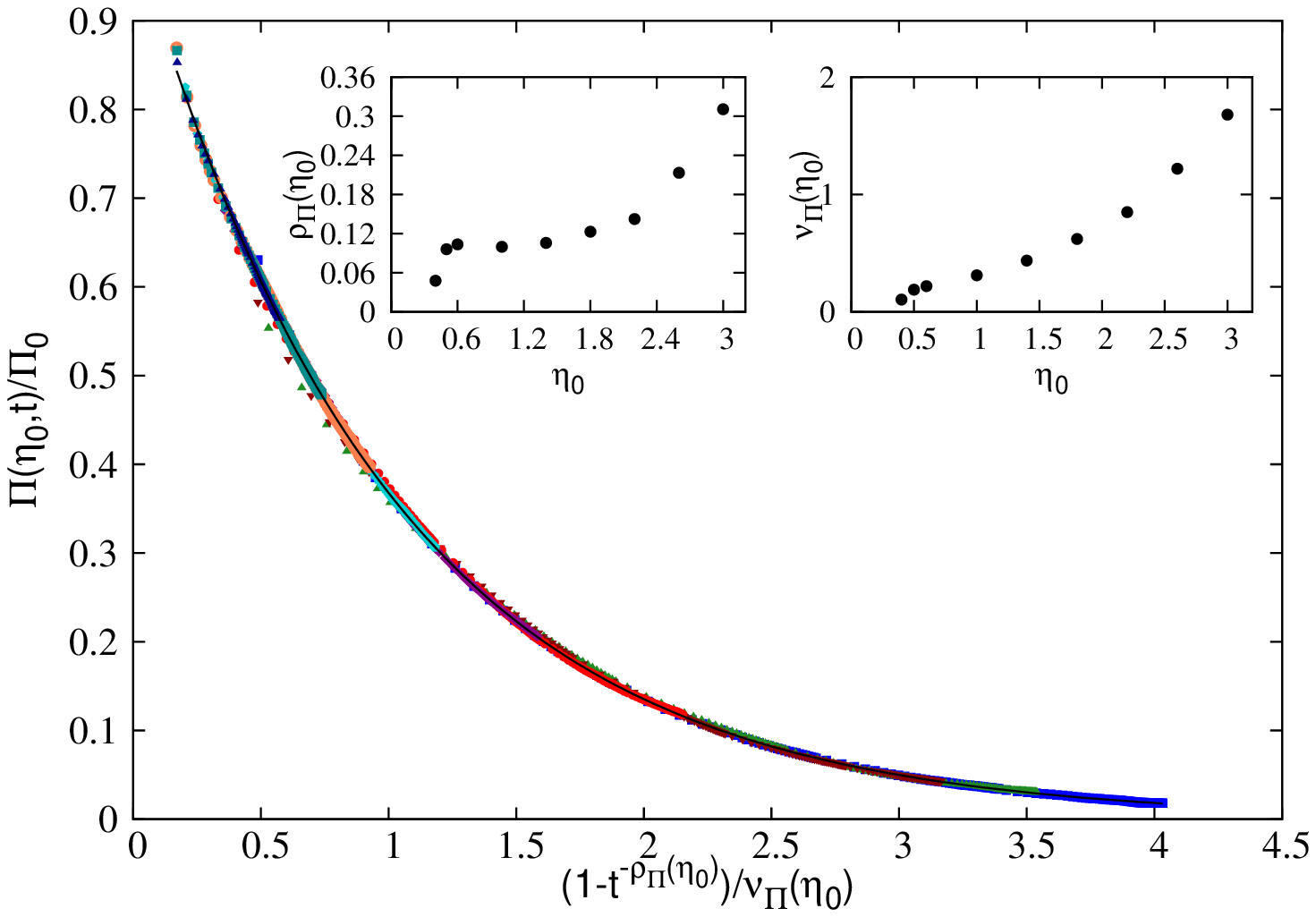}
  \caption{(Color online) Plot of the data collapse of the function 
    $\Pi(\eta_0,t)$.
    The inset shows the variation of $\rho_\Pi(\eta_0)$ and
    $\nu_\Pi(\eta_0)$ against $\eta_0$.}
  \label{fig4}
\end{figure}

\begin{figure}[h!]
  \centering
  \includegraphics[scale=0.30,angle=-90]{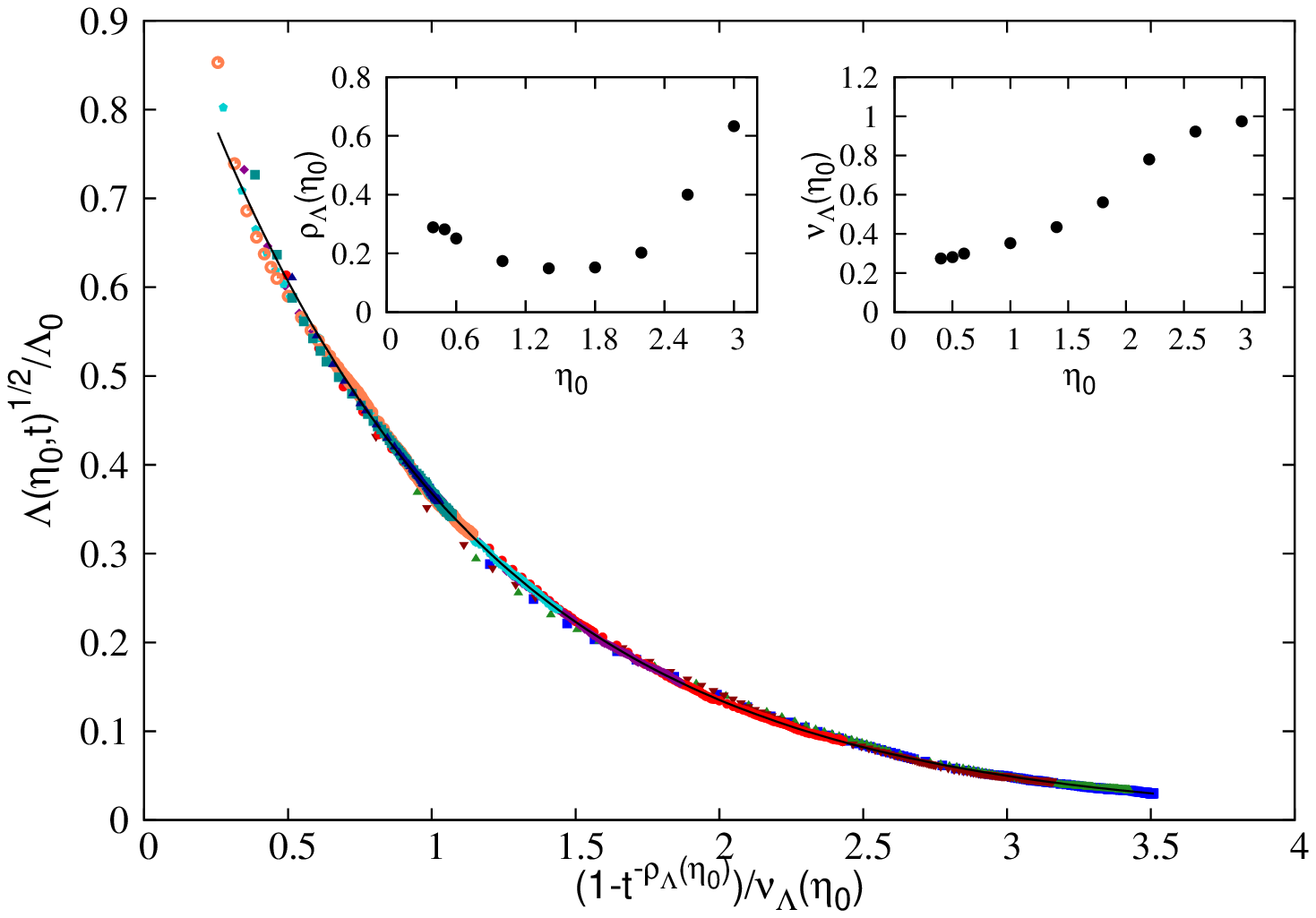}
  \caption{(Color online) Plot of the data collapse of the function
    $\Lambda(\eta_0,t)^{1/2}$. The inset shows the variation of
    $\rho_\Lambda(\eta_0)$ and $\nu_\Lambda(\eta_0)$ against
    $\eta_0$.}
  \label{fig5}
\end{figure}

\begin{figure}[h!]
  \centering
  \includegraphics[scale=0.30,angle=-90]{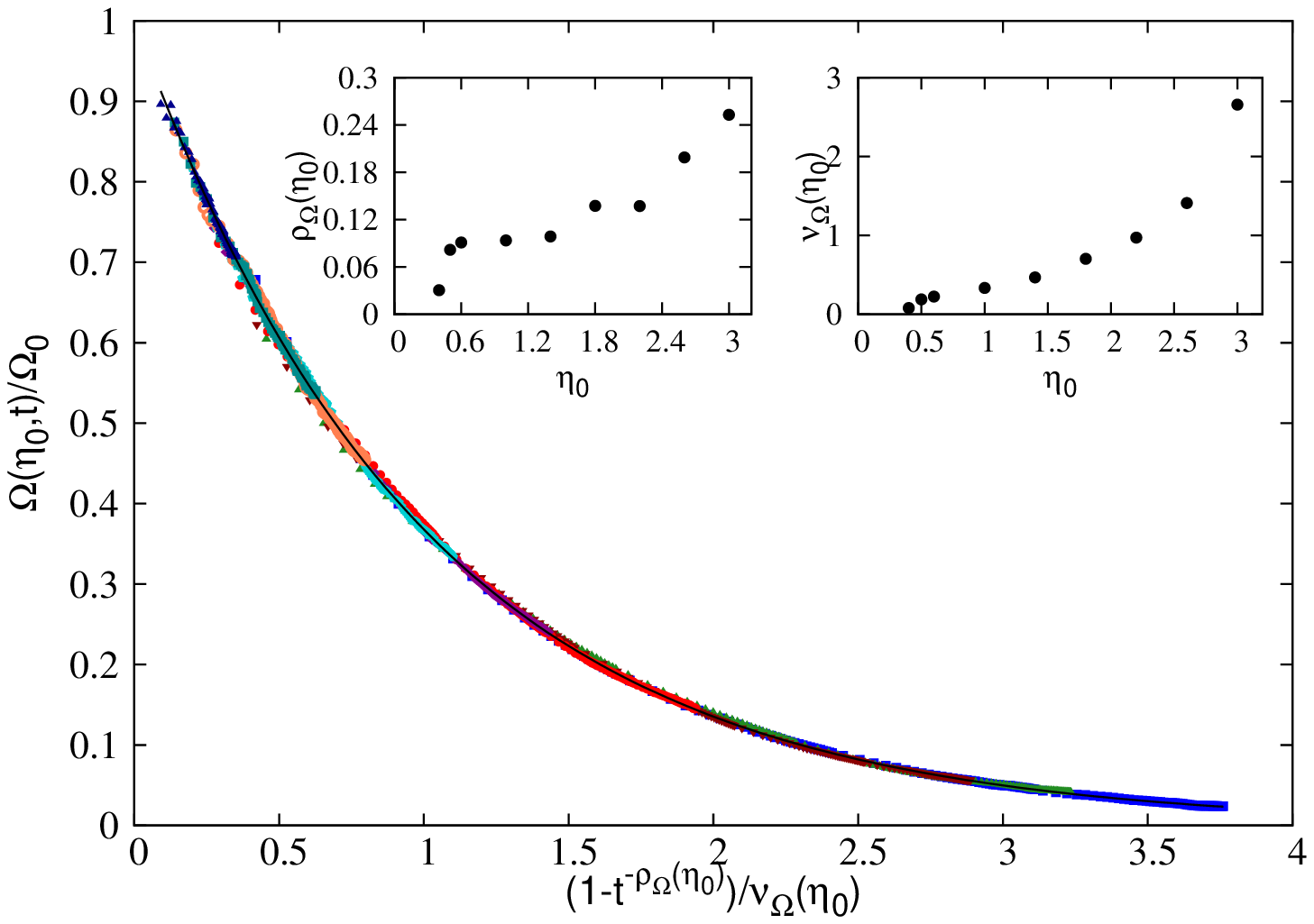}
  \caption{(Color online) Plot of the data collapse of the function  
   $\Omega(\eta_0,t)$. The inset shows the variation of $\rho(\eta_0)$ and
    $\nu(\eta_0)$ against $\eta_0$.}
  \label{fig6}
\end{figure}

We begin our analysis with the idea put forward by Imry and Ma
\cite{im}. They argued that if one reverses the spins within a domain of 
linear size $R$, the energy cost $E_{\tt{ex}}$ is proportional to the 
domain wall area, i.e., $E_{\tt{ex}} \propto JR^{d-1}$ where $d$ is the spatial
dimension. This energy increase has to be compared with the energy
gain from the interaction with the random fields. The central limit
theorem tells that the mean square random field energy $E_{\tt{RF}}^2$
inside a region of volume $R^d$ is $\sim \eta_0^2 R^d$. The total
energy involved in the creation of a domain of linear size
$R$ is therefore
\be
\label{eqn2}
   E(R) \approx J R^{d-1} - \eta_0 R^{d/2}
\ee
The first term of Eq. (\ref{eqn2}) represents the contribution due to
the boundary of the domain with linear size $R(t)$. The second term
represents the contribution due to the fluctuations of the random
fields in the bulk of the domain of linear size $R(t)$. 
On the basis of the above argument, Imry and Ma concluded that the 
lower critical dimensionality (LCD) of the RFIM is two. This argument 
is based on domains having flat interfaces. The question arises if this 
argument would hold even in presence of rough interfaces with fluctuating 
curvature. To address this concern, Binder \cite{bind}
reformulated the problem in terms of the interfaces and had shown that
the interface roughness is negligible if $d > 2$, confirming 
that the LCD of RFIM is two. 
However, Binder shows that the domain wall energy has a logarithmic 
correction, which introduces a break up length scale '
$L_b \sim {\tt exp}(A[J/\eta_0]^2)$ below which Imry-Ma argument is valid. 
Well defined interfaces of domains are meaningful for length scales
less than $L_b$. The system size considered here is below the $L_b$ and 
therefore it makes sense to consider Eq. (\ref{eqn2}) as the starting point 
of our analysis, although Eq. (\ref{eqn2}) disregards the logarithmic 
correction. Taking a cue from Eq. (\ref{eqn2}), the
surface energy of all the domains $E_{\tt{ex}}^{\tt{t}} \sim
J\Pi(\eta_0,t)$ and the mean square bulk energy contained in
  all the domains of the system due to random fields
${E_{\tt{RF}}^{\tt{t}}}^2 \sim \frac{{\eta_0}^2{L^d}}{{\Lambda(\eta_0,t)}}$,
as the density of domains is inversely related to their characteristic
volume. Thus the energy density ($\epsilon = E^t/L^d$) involved in
the creation of all the domains in the system is given by
\be
\label{eqn2a}
  \epsilon(\eta_0,t) \approx \frac{J\Pi(\eta_0,t)}{L^{d}} -
  \frac{\eta_0}{{\Lambda(\eta_0,t)}^{1/2}}
\ee
As $L \rightarrow \infty$, the surface energy
contribution vanishes. This is true for any growing
volume. For finite system size, the contribution 
from the surface energy term cannot be neglected. So the
variation of either $\Pi(\eta_0,t)$ or $\Lambda(\eta_0,t)^{1/2}$ with time
will govern the coarsening of the system. The
log-log plot of $\Pi(\eta_0,t)$ and $\Lambda(\eta_0,t)^{1/2}$ against time
are shown in Fig. \ref{fig1} and \ref{fig2} respectively. 
It is evident from Eq. (\ref{eqn2a}) that the coarsening of the
system energetically favours the minimization of the total 
length of the interfaces and also the decrease in the number of the
domains. This is observed in Fig. \ref{fig1} and \ref{fig2} respectively. In view of the above
discussions, we can redefine the coarsening as simply the minimization
of $\epsilon(\eta_0,t)$ and the kinetic coarsening will be
characterized by the scaling behaviour of 
$\Pi(\eta_0,t)$ and $\Lambda(\eta_0,t)^{1/2}$. As time flows,
small domains coalesce to form relatively larger domains and from
Eq. (\ref{eqn2}), it is clear that a domain with a typical average size $R$
should grow in such a way that the total length of all the interfaces of 
all the domains present in the system  shrinks in order to
minimize the energy of the system. The domain with largest mass should
grow in the same fashion during its dynamical evolution.
Therefore $\Omega(\eta_0,t)$ is expected to exhibit similar behaviour 
as that of $\Pi(\eta_0,t)$. Log-log plot of $\Omega(\eta_0,t)$ against 
time is shown in Fig. \ref{fig3}.
In this context, Sepp\"al\"a and Alava \cite{sa} showed
that below a critical random field strength, the largest
domain spans the system and the $2D$ RFIM shows a percolation transition. 
This was supported by some later studies \cite{lkfi,kpi}. 
We also check that below a critical random field strength $\eta_c$, the
largest cluster is a spanning one with a fractal dimension 
$1.89 \pm 0.02$. Above $\eta_c$, the largest cluster is finite. The
value of $\eta_c$ of course depends on temperature.
In a recent article \cite{sm} we also reported that below a critical
random field strength, the $2D$ RFIM exhibits long range order (LRO).
We now provide the theory of kinetic coarsening.  
In general, we denote by $\Psi(\eta_0,t)$ the three quantities. 
A careful observation of the graphs suggests that the initial and  asymptotic
behaviour of the generalized function $\Psi(\eta_0,t)$ is given by
\bea
\label{eqn3}
\Psi(\eta_0,t) \rightarrow \Psi_0 \,\,(\text{constant}), \,\,\, \text{as}~~ t
\rightarrow 1 \n \\
{\text{and}}~~ \Psi(\eta_0,t)\rightarrow \Psi_0 e^{ - 1/\nu(\eta_0)},
\,\,\, \nu(\eta_0) > 0, ~~\text{as}~~ t \rightarrow \infty
\eea
where $\nu(\eta_0)$ is a disorder-dependent scaling exponent. The
decay rate of the function $\Psi(\eta_0,t)$ at any time step for a
fixed $\eta_0$ should depend on the following factors: first, on the 
value of the function itself at this time step. Secondly, from the nature
of the variation of the functions, it is
evident that the rate of decay of the function $\Psi(\eta_0,t)$ also
depends on the particular time step. As the time flows, the rate of
decay of $\Psi(\eta_0,t)$ slows down and this dependence is taken as
a power law decay. In addition to these factors,
another $\eta_0$-dependent parameter should be there for
controlling the decay rate of $\Psi(\eta_0,t)$. This parameter considers the
wandering of the interfaces in presence of the random fields. Thus the
decay rate of $\Psi(\eta_0,t)$ is given by
\be
\frac{d\Psi}{dt} \sim - a(\eta_0)\Psi t^{-\mu(\eta_0)},~~~ \mu(\eta_0)>0
\label{eqn4}
\ee
$a(\eta_0)$ is a disorder-dependent parameter. Integrating,
\be
\Psi(\eta_0,t) = \Psi_0 {\tt
  exp}\left[\frac{t^{-(\mu(\eta_0)-1)}}{\nu(\eta_0)} +
  k(\eta_0)\right]
\label{eqn5}
\ee
where $k(\eta_0)$ is a constant of integration and $\nu(\eta_0) =
\frac{\mu(\eta_0)-1}{a(\eta_0)}$. Now from (\ref{eqn3}) as $t \to 1$,
$\Psi(\eta_0,t) \to \Psi_0$ which gives $k(\eta_0) = -1/\nu(\eta_0)$
and as $t \to \infty$, $\Psi(\eta_0,t)\rightarrow \Psi_0 e^{ -
  1/\nu(\eta_0)}$ which gives $\mu(\eta_0) > 1$. Thus the functional
form of $\Psi(\eta_0,t)$ is given by
\be
\Psi(\eta_0,t) = \Psi_0 {\tt
  exp}\left[-\frac{1-t^{-\rho(\eta_0)}}{\nu(\eta_0)}\right]
\label{eqn6}
\ee
where $\rho(\eta_0) = \mu(\eta_0)-1 > 0$. The scaling behaviour
(\ref{eqn6}) of the functions characterizing the kinetic coarsening
shows an universal nature with two disorder-dependent exponents
$\rho(\eta_0)$ and $\nu(\eta_0)$. The validity of the
the scaling relation (\ref{eqn6}) can be confirmed from the plots of
the data collapse of the functions $\Pi(\eta_0,t)$,
$\Lambda(\eta_0,t)^{1/2}$ and $\Omega(\eta_0,t)$ with the corresponding
exponents $(\rho_\Pi,\nu_\Pi), (\rho_\Lambda,\nu_\Lambda) ~\text{and}~
(\rho_\Omega,\nu_\Omega)$ \cite{supm}. The plots are shown in Fig. \ref{fig4},
\ref{fig5} and \ref{fig6} respectively.   
From scaling relation (\ref{eqn6}) the initial time behaviour of 
$\Psi(\eta_0,t)$ is given by
\be
  \Psi(\eta_0,t) = \Psi_0 t^{-\rho(\eta_0)/\nu(\eta_0)} ~~\text{for}~~
  t \ll \text{exp} (1/\rho)
\label{eqn7}
\ee
Thus the function $\Psi(\eta_0,t)$ shows a power law decay with
the exponent $\frac{\rho(\eta_0)}{\nu(\eta_0)}$ till the
characteristic time scale $t_\times \sim e^{1/\rho(\eta_0)}$. This
initial linear behaviour in log-scale is observed from 
Fig.\ref{fig1}, \ref{fig2} and \ref{fig3} respectively. 
It is observed from the insets of the Fig. \ref{fig4}, \ref{fig5}
and \ref{fig6} ,where the variation of $\rho(\eta_0)$ and $\nu(\eta_0)$ 
against $\eta_0$ is shown, that as $\eta_0 \to 0$,
$\rho(\eta_0) \to 0 ~\text{and}~ \nu(\eta_0) \to 0 ~\text{with}~
\frac{\rho}{\nu} ~\text{finite}$, which implies that the power law
decay continues for longer time as $\eta_0 \to 0$. This behaviour is
quite expected because for weak random field strength, the
decay of $\Psi(\eta_0,t)$ is dominated by the exchange interaction.
$\Psi(\eta_0,t)$ asymptotically approaches the value $\Psi_0
e^{-1/\nu(\eta_0)}$. Physically it means that as $t \to \infty$, the
domains cease to grow.
The dynamic behaviour of the average linear domain size can also be
predicted from the scaling relation (\ref{eqn6}). The
typical average linear size  $R(\eta_0,t)$ is given by
\begin{equation}
  \begin{split}
     R(\eta_0,t)^d & \sim \frac{L^d}{\Lambda(\eta_0,t)} \\
     R(\eta_0,t)   & \sim \Lambda(\eta_0,t)^{-1/2} ~~~\text{for}~~ d=2
  \end{split}
  \label{eqn8}
\end{equation}
From simple calculations, the initial time, late time and the
asymptotic nature of $R(\eta_0,t)$ are obtained as
\begin{equation}
  \begin{split}
    R(\eta_0,t)  & \sim t^{\rho_\Lambda/\nu_\Lambda} ~~\text{for}~~ t
    \ll e^{1/\rho_\Lambda} \\
                 & \sim (\ln t)^{1/\nu_\Lambda} ~~\text{for}~~
    e^{1/\rho_\Lambda} \ll t \ll \infty  \\
                 & \sim e^{1/\nu_\Lambda} ~~\text{for}~~ t \to \infty
  \end{split}
  \label{eqn9}
\end{equation}
We interpret the ratio 
$\frac{\nu_\Lambda(\eta_0)}{\rho_\Lambda(\eta_0)}$ as the non universal 
dynamic exponent $z(\eta_0)$ corresponding to the
early time power law growth of $R(\eta_0,t)$. Also, the barrier
exponent for the late time regime is interpreted as $\nu_\Lambda(\eta_0)$. 
Thus the scaling relation (\ref{eqn6})
successfully reproduces the recent claims \cite{puri,np,mg,ibk} that
the growing length scale $R(\eta_0,t)$ shows a dynamic crossover from
a pre-asymptotic algebraic growth to asymptotic slow logarithmic
growth. Another essential feature corresponding to the
growth of $R(\eta_0,t)$ is contained in the proposed scaling law. At $t
\to \infty$, the value of $R(\eta_0,t)$ approaches to 
$e^{1/\nu_\Lambda(\eta_0)}$. This avoids the asymptotic divergence of
$R(\eta_0,t)$. Thus, the behaviour of the average linear
domain size $R(\eta_0,t)$ in the entire time regime can physically be
explained with the help of the scaling relation (\ref{eqn6}).
In view of the above discussion, $\nu(\eta_0)$ is interpreted as
follows. The scaling relation (\ref{eqn6}) shows
that as $\nu(\eta_0) \to 0$, $\Psi(\eta_0,t)
\to 0$ for $t \to \infty$. It means that the pinning interaction starts 
dominating as $\nu(\eta_0)$ increases. Thus, $\nu(\eta_0)$ is responsible
for the stiffness of the domain wall. This is also obvious from 
the late time dynamics of $R(\eta_0,t)$ which is governed by the exponent
$\nu(\eta_0)$ only (see Eq.(\ref{eqn9})). The domain wall
would become stiffer with the increase of $\nu(\eta_0)$
As $t \to 1$, $\Psi(\eta_0,t)$
reaches a fixed value $\Psi_0$, independent of $\eta_0$.
It is to be noted that the scaling relation (\ref{eqn6}) suggests that
as $t \to \infty$,
$\Psi(\eta_0,t)/\Psi_0 = {\tt exp} (-1/\nu)$. So at
$t \to \infty$, the number of domains relative to their
initial value converges to a well-defined value and the
ratio is a measure of the entropy of the system \cite{im}.
%Again if $\eta_0$ is known, the value of $\nu (\eta_0)$ 
%can be found from the insets of Fig. 3, 4 and 5.
Thus if the ratio is known, the value of $\nu (\eta_0)$
can be determined and from the insets of Fig. \ref{fig4}, 
\ref{fig5} and \ref{fig6}, the value of $\eta_0$ may be found 
corresponding to a particular $\nu (\eta_0)$.
Therefore the infinite time limit of the scaling relation
converges to a well-defined thermodynamic quantity that
would fix the value of $\eta_0$.

We end this article with a few comments.
Although we present results for a particular temperature and for a
particular system size, we check that 
the same scaling relation holds good for other temperatures 
and other system sizes as well.
%In this sense, the scaling relation turns out to be universal.
However, the system size has to be below the break up length 
scale and the temperature should not be so high that the 
thermal fluctuations become relevant. 
%The exponents in the insets
%of Fig. \ref{fig4},\ref{fig5} and \ref{fig6} appear to be very
%small for weak random field strength. This might be due to the 
%finite system size.
We would also like to point out that the Hamiltonian given by Eq.(\ref{eqn1})
depends on $J$, $\eta_0$ and $T$, or more precisely on the ratio
$J/T$ and $\eta_0/T$. $J/T$ being fixed in the present work, the quantities
of our interest depend on $\eta_0/T$ only. This means irrespective of any 
particular value of $\eta_0$ and $T$, the ratio of $\eta_0$ and $T$ would govern
the coarsening of the system. This adds generality to the scaling relation (\ref{eqn6}).
Although we arrived
at the scaling relation for the 2D RFIM, this relation also corroborates 
the recent claim of a possible crossover from a early time power law growth 
to a late-time logarithmic growth in Ising model with random coupling and
random dilution \cite{puri2,puri3}. Certainly many more simulations on different systems
are required to confirm the generic nature of the scaling relation. (work
in this direction is in progress)

One of the authors (SS) acknowledges support from the UGC Dr. D. S.
Kothari Postdoctoral Fellowship under grant
No. F.4-2/2006(BSR)/13-416/2011(BSR). SS also thanks Heiko Rieger for
many useful discussions. The authors acknowledge gratefully  
anonymous referees for a number of critical comments.

\end{document}